\documentclass[twocolumn]{aastex701}
\usepackage{amsmath, dsfont, hyperref}

\newcommand{\Lagr}{\mathcal{L}}
\newcommand{\Lpost}{\Lagr_\mathrm{post}}

\newcommand{\SNR}{\mathrm{SNR}}
\newcommand{\trho}{\tilde{\rho}}
\newcommand{\thetapost}{\hat{\theta}_\mathrm{post}}
\newcommand{\Sigmapost}{\hat{\Sigma}_\mathrm{post}}

\begin{document}

\title{An Automated Probabilistic Asteroid Prediscovery Pipeline}

\author[0009-0000-0521-5965, sname=Li]{Sage Li}
\affiliation{Physics Division, Lawrence Livermore National Laboratory, 7000 East Avenue, Livermore CA, 94550, USA}
\affiliation{Georgia Institute of Technology, 225 North Avenue, Atlanta GA, 30332, USA}
\email[show]{sli774@gatech.edu}  

\author[0000-0002-1269-6047, sname=Geringer-Sameth]{Alex Geringer-Sameth}
\affiliation{Physics Division, Lawrence Livermore National Laboratory, 7000 East Avenue, Livermore CA, 94550, USA}
\email[show]{geringersame1@llnl.gov}

\author[0000-0003-2632-572X, sname=Golovich]{Nathan Golovich}
\affiliation{Physics Division, Lawrence Livermore National Laboratory, 7000 East Avenue, Livermore CA, 94550, USA}
\email[show]{golovich1@llnl.gov}

\begin{abstract}
We present an automated and probabilistic method to make prediscovery detections of near-Earth asteroids (NEAs) in archival survey images, with the goal of reducing orbital uncertainty immediately after discovery. We refit the Minor Planet Center's astrometry and propagate the full six-parameter covariance to survey epochs to define search regions. We build low-threshold source catalogs for viable images and evaluate every detected source in a search region as a candidate prediscovery. We eliminate false positives by refitting a new orbit to each candidate and probabilistically linking detections across images using a likelihood ratio. Applied to Zwicky Transient Facility (ZTF) imaging, we identify approximately 3000 recently discovered NEAs with prediscovery potential, including a doubling of the observational arc for about 500. We use archival ZTF imaging to make prediscovery detections of the potentially hazardous asteroid 2021~DG1, extending its arc by 2.5~yr and reducing future apparition sky plane uncertainty from many degrees to arcseconds. We also recover 2025~FU24 nearly 7~yr before its first known observation, when its sky plane uncertainty covers hundreds of square degrees across thousands of ZTF images. The method is survey agnostic and scalable, enabling rapid orbit refinement for new discoveries from Rubin, NEO Surveyor, and NEOMIR.
\end{abstract}

\section{Introduction}

The discovery, tracking, and characterization of near-Earth asteroids (NEAs) are central to planetary defense. Accurate orbit determination is fundamental for assessing impact probabilities, constraining long-term dynamical evolution, and planning follow-up campaigns. The mathematical framework for orbit fitting originates with Gauss, but the modern era of asteroid orbit determination has been shaped by advances in statistical methods for uncertainty propagation and impact monitoring \citep[e.g.][]{milani2010theory}. In particular, the development of statistical ranging \citep{2001Icar..154..412V}, Monte Carlo sampling techniques \citep{milani2005nonlinear}, and systematic impact monitoring pipelines such as CLOMON2 and Sentry \citep{farnocchia2015systematic} have enabled robust evaluation of orbital uncertainties even in the short-arc regime.  

A critical opportunity lies in the identification of \emph{prediscovery} detections (or ``precoveries''), where archival survey images contain serendipitous detections of an asteroid prior to its official discovery. Such data extend observational arcs, often by years, dramatically improving orbital accuracy and impact risk assessments. Systematic prediscovery efforts have been pursued for decades, including early work on recovering NEAs \citep{boattini2000recovery}, the Arcetri Near-Earth Object Precovery Program \citep{2001A&A...375..293B}, and large-scale archival mining initiatives such as EURONEAR and its Mega-Precovery extension \citep{2009AN....330..698V,2013AN....334..718V}. More recent projects have used citizen science and virtual observatory tools \citep{2013hsa7.conf..995R,2014AN....335..142S}, and case studies have demonstrated the scientific value of precoveries, such as prediscovery imaging of the disrupted asteroid P/2010~A2 \citep{2011AJ....142...28J} and the temporary Earth satellite 2020~CD$_3$ \citep{2021ApJ...913L...6N}. Contemporary work continues to refine search strategies, from improved statistical modeling of uncertainty regions \citep{2024A&A...689A..49V} to the mining of modern wide-field archives \citep{2023A&A...673A..93S}. The B612 Foundation has also developed a web-based tool for prediscovery,\footnote{\url{https://b612.ai/adam-platform/precovery/}} which propagates input asteroids to catalogs from public surveys. 

We present an automated prediscovery pipeline that combines orbit fitting to the Minor Planet Center's (MPC) astrometry, propagation of the orbital uncertainty distribution into archival and ongoing wide-field surveys, and a linking step to control false positives. By constructing probabilistic sky maps instead of single-point ephemerides, the pipeline enables statistically robust searches that account for both astrometric uncertainty and the sensitivity of archival survey images. This approach is survey agnostic (though it requires access to survey images or low-threshold source catalogs) and is applicable to current and future wide-field facilities such as Pan-STARRS \citep{chambers2016panstarrs}, ZTF \citep{masci2019ztf} and the Rubin Observatory \citep{ivezic2019lsst}.

The discovery rate of near-Earth objects (NEOs) is set to rapidly increase, primarily due to the Rubin Observatory, NEO Surveyor \citep{2023PSJ.....4..224M}, and NEOMIR \citep{2024EPSC...17..882L}. These newly discovered asteroids will potentially have nonzero impact probabilities and require rapid prediscovery searches. Our prediscovery pipeline is designed to carry out this task on wide-field surveys extending years into the past from the discovery date. The resulting prediscovery identifications extend asteroid arcs, improve orbit determinations, and ultimately strengthen planetary defense capabilities. In this paper, we demonstrate the method with difference imaging produced by ZTF over approximately 6~yr from 2018 to 2024.

The remainder of this paper is organized as follows. Section~\ref{sec:data} outlines the ZTF survey data used in this paper.
In Section~\ref{sec:Methods}, we describe our pipeline in detail, including the orbit fitting procedure, numerical propagation, and our candidate detection and statistical linking algorithms.  
In Section~\ref{sec:results}, we quantify the prediscovery potential of ZTF and demonstrate successful prediscovery of the NEAs 2021~DG1 and 2025~FU24 in archival data. Finally, Section~\ref{sec:conclusions} summarizes our findings, outlines future directions for scaling our prediscovery pipeline, and discusses prospects for application to other survey datasets.

\section{Data\label{sec:data}}
Our prediscovery system is a general method that can be implemented with any sky survey (or data from multiple surveys). For concreteness, the remainder of this paper considers an application to the Zwicky Transient Facility \citep{2019PASP..131a8002B}. 

ZTF is large-etendue survey observing the entire sky north of declination $-30~\mathrm{deg}$. Its coverage is such that nearly all NEAs are present in survey images (often, of course, at magnitudes too faint to detect). We have acquired all $r$ and $g$-band difference imaging \citep{masci2019ztf,https://doi.org/10.26131/irsa539} from March 20, 2018 to February 29, 2024. Our data set consists of 849,750 exposures, each 30 seconds in duration and split into 64 ccd-quadrant FITS files. Each quadrant is a $3072\times3080$-pixel image, corresponding to a sky area of about $0.75~\mathrm{deg}^2$. The $5\sigma$ limiting magnitude for point source detection is around 20.6 (similar in $r$ and $g$). The median seeing full width at half max is about 2.0 pixels for $r$ and 2.2 for $g$.

\section{Methods}
\label{sec:Methods}
\begin{figure}[b!]
    \centering
    \includegraphics[width=0.9\linewidth]{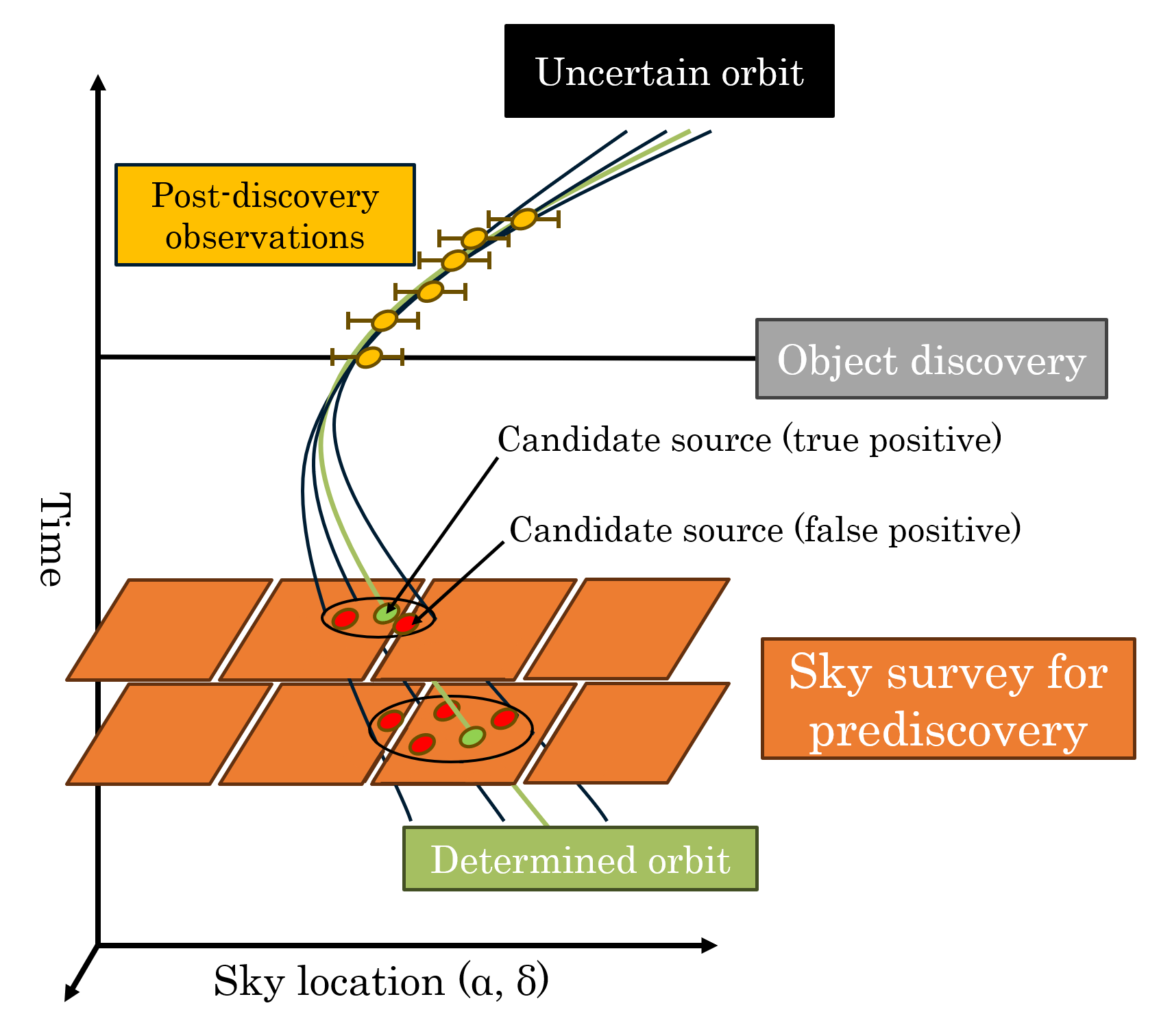}
    \caption{A visualization of the prediscovery algorithm. The orbital uncertainty from postdiscovery observations is propagated back in time to the survey to identify images and search regions within those images (black ellipses). Sources detected in the search regions are candidate prediscoveries. Each candidate source corresponds to a trial orbit, which is propagated to the other images to look for coincident detections. A true prediscovery will result in multiple detections along a single physical orbit.\label{fig:overview}}
\end{figure}

The following subsections describe the processing steps of our pipeline in detail. Figure \ref{fig:overview} shows a schematic representation. 

\subsection{Fitting procedure}
\label{subsec:fitting_and_loss}
The method begins with the set of $N$ observations listed in the MPC\footnote{\url{https://www.minorplanetcenter.net}} for a given NEA, which we gather using the MPC API\footnote{\url{https://minorplanetcenter.net/mpcops/documentation/observations-api}}. We call these the ``postdiscovery'' data.
Since the MPC does not provide the full covariance for its orbit determinations, we implement a least-squares fitting procedure to obtain a best-fit orbit and covariance. Our ephemeris prediction uses ASSIST \citep{assist} for orbit propagation and Astropy \citep{2022ApJ...935..167A} to calculate the position and velocity of an observatory at the time of observation. ASSIST is based on the IAS15 integrator \citep{2015MNRAS.446.1424R} and the JPL DE441 solar system ephemeris model \citep{Park2021DE440}. Fitting is done in the six Keplerian orbital elements at an epoch set to the midpoint of the ZTF survey.

For a set of Keplerian elements $\theta$, each observation $x_i$ (either RA or Dec) contributes a $\chi^2$ term
\begin{equation}
\chi^2(x_i \mid \theta, \sigma_i) = \left( \frac{x_{i} - f(\theta, t_i)}{\sigma_i} \right)^2, \label{eq:chi2factor}
\end{equation}
where $f(\theta, t_i)$ is the ephemeris prediction at the time of the observation $t_i$ and $\sigma_i$ is the associated error in $x_i$. For MPC observations without a reported uncertainty we assign conservative 2~arcsec errors independently in the RA and Dec directions. Correlations between RA and Dec uncertainties are typically not reported, but it would be straightforward to include them in the $\chi^2$ term.

Our initial fits to MPC data resulted in solutions with poor goodness of fit, likely due to biases in reported observations. We therefore adopted a modified $\chi^2$ loss to discount outliers. Each $\chi^2_i$ term is replaced with $\rho(\chi^2_i, c)$, where $\rho(z,c)$ is the Huber function
\citep{huber},
\begin{equation} \nonumber
    \rho(z, c) = 
    \begin{cases}
    z & \text{if } z/c^2 \leq 1, \\
    c^2(\sqrt{z/c^2} - 1) & \text{if } z/c^2 > 1.
    \end{cases}
\end{equation}
The Huber terms behave like the usual $\chi^2$ (i.e., as mean squared error) as long as the residuals between data and model are less than $c$ standard deviations ($\lvert x_i - f(\theta,t_i) \rvert < c \sigma_i$). Above this threshold, a measurement is treated as an outlier and the loss only increases as the absolute error rather than as the squared error. The influence of outlier points on the fit is reduced in the same way as when replacing a mean estimate with a median. The Huber factor attenuates biased observations while maintaining the effectiveness of ordinary least squares. The total loss function for fitting to the postdiscovery data is
\begin{equation}
\Lpost(\theta, c) = \sum_{i=1}^{2N}\rho\left(\chi^2(x_i \mid \theta, \sigma_i), \,c\right).
\label{eq:Lpost}
\end{equation} 

We use the trust-region reflective minimization algorithm implemented in SciPy \citep{2020SciPy-NMeth}, returning a best-fit set of orbital elements $\thetapost$ and a Fisher information-based estimate of the orbital element covariance $\Sigmapost$. Through manual testing, we find a choice of $c=1$ to yield the desired balance of convergence time and robustness (from now on we drop the argument $c$).

We query JPL's Horizons system \citep{NASAJPLHorizons,1996DPS....28.2504G} to obtain a set of orbital elements at our epoch to use as an initial guess for the minimization algorithm.

\subsection{Identification of potential prediscovery images\label{sec:imageidentification}}

The postdiscovery uncertainty in orbital elements can be projected onto the sky at the time of each survey image to determine the images and regions within those images where the asteroid may likely be found. 

At a given confidence level, the postdiscovery fit has constrained the orbit to lie within a 6-d ellipsoid in the space of orbital elements. We map out the boundary of this region by drawing 20{,}000 orbital element samples from an approximate isometric $\chi^2$ surface corresponding to the 99.99\% confidence level. Specifically, we set $\chi^2_q$ to the $q=0.9999$ quantile of a $\chi^2$ distribution with 6 degrees of freedom and then sample orbits $\theta$ satisfying ${(\theta - \thetapost)^T \Sigmapost^{-1} (\theta-\thetapost) = \chi^2_q}$, where $\thetapost$ are the best-fit orbital elements obtained in Section~\ref{subsec:fitting_and_loss}. This is done using the Cholesky composition $\Sigmapost = LL^T$ and generating sample orbits via $\theta = \thetapost + \sqrt{\chi^2_q} \,L z$, where $z$ is sampled on the 6-d unit sphere using \texttt{scipy.stats.uniform\_direction(dim=6)}. As we care about the boundary of the allowed region rather than its interior, this procedure is more efficient than simply sampling from a 6-d Gaussian distribution centered on $\thetapost$ with covariance $\Sigmapost$.

Then, for each sample orbit, we propagate its predicted trajectory through the survey to find the images it intersects using a fast, scalable search algorithm. Images taken during the postdiscovery observational arc are disregarded.

NEAs are typically visible only for short time periods near close approaches. To avoid wasted effort, we only consider images in which the asteroid's predicted flux is plausibly close to the detection threshold. We implement the H-G model \citep{1989aste.conf..524B} to estimate the light curve based on the postdiscovery orbit fit and reject images in which the asteroid's flux is below the 3$\sigma$ point source detection threshold for the image (this threshold is softened by a magnitude to allow for uncertainties in the postdiscovery measurements, rotational variations, and color correction; $H$ and $G$ are set to their MPC reported values, including the default $G=0.15$ assumed for asteroids with no measured $G$).

The surviving images plausibly contain the asteroid at magnitudes bright enough to detect. For each intersection, we use the image's World Coordinate System solution \citep{2002A&A...395.1077C} to get the pixel coordinates of the sampled orbits.

\begin{figure*}[t]
    \centering
    \includegraphics[width=0.8\linewidth]{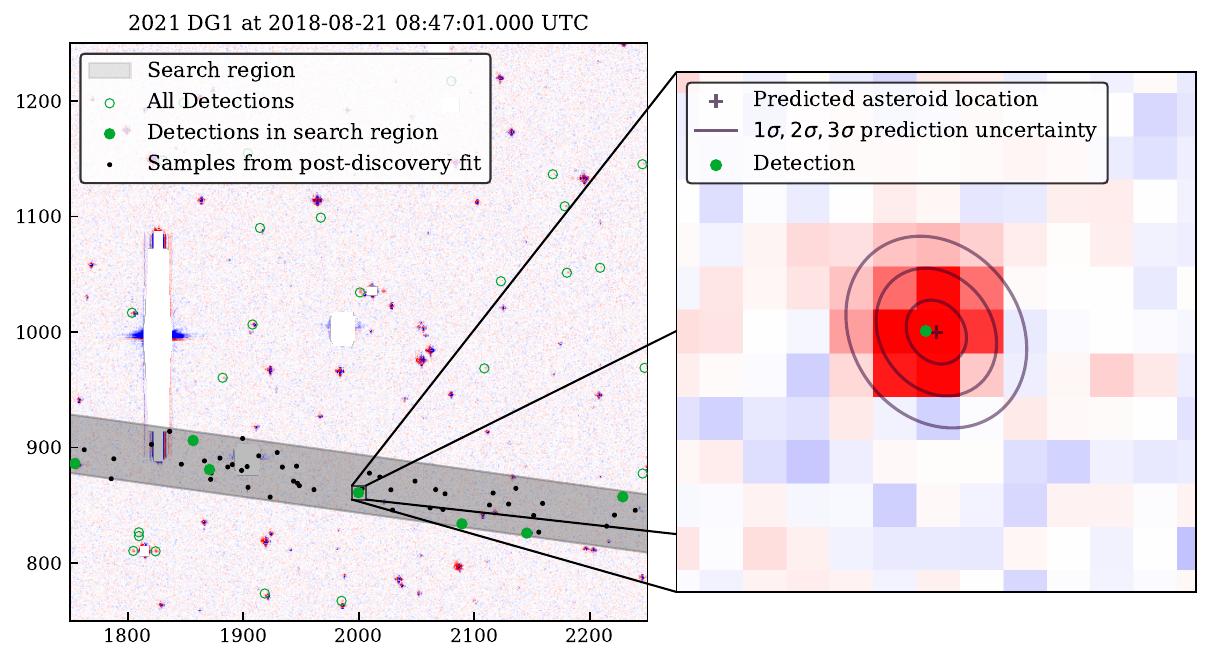}
    \caption{Prediscovery of the near-Earth asteroid 2021~DG1 in a ZTF image from August 2018, 2.5~years before its discovery date. For clarity, the left image is zoomed to a $500 \times 500$ pixel subregion of the full $3072\times 3080$ difference image. Black points are locations of sample orbits consistent with the postdiscovery observations. They are used to construct a convex hull (gray region) which forms the search region for this image. Image sources detected above $4\sigma$ significance are shown as green circles (filled if they are in the search region, empty if outside it). A detected source in another image is considered to be a candidate prediscovery and the resulting trial orbit is propagated to this image. The inset shows the predicted location ($+$) and 1$\sigma$, 2$\sigma$, and 3$\sigma$ confidence regions (black contours) for this updated orbit. The trial orbit is consistent with the location of a source  detected in this image (green point), contributing greatly to this particular orbit's likelihood ratio.
    \label{fig:ztfimage}}
\end{figure*}

To define a search region within an image, we construct a convex hull\footnote{We use the SciPy function \texttt{scipy.spatial.ConvexHull}.} \citep{10.1145/235815.235821}, which is a minimum convex region enclosing the coordinates of all the orbit samples. Our testing shows the sample points to always occupy a convex region when restricted to a single image. However, for cases when the image date is long before the postdiscovery data, the sampled orbits can land far beyond the image borders in a banana-shaped region, making the convex hull much larger than it needs to be. To avoid this, we remove sample orbits that land beyond a 500 pixel buffer outside the image boundary. An illustration of the convex hull construction is shown in Figure~\ref{fig:ztfimage}.

\subsection{Source detection\label{sec:source_detection}}
The next step is to perform source detection on the identified images. Sources within the search regions are considered candidate detections of the asteroid.

For each image we construct a catalog of sources detected within the search region using the DAOFIND algorithm \citep{1987PASP...99..191S} implemented in the \texttt{DAOStarFinder} function\footnote{The image noise level is approximated with \texttt{astropy.stats.sigma\_clipped\_stats} and subtracted before passing the image to \texttt{DAOStarFinder}.} in the Photutils package \citep{larry_bradley_2025_14889440}. DAOFIND finds sources that exceed a specified significance level. At this step of the analysis, the priority is that the asteroid makes it into the catalog and we are relatively unconcerned about including spurious sources (false positives are handled in the next step). We set the detection threshold at $4\sigma$, but reducing this parameter is likely to be beneficial. The PSF FWHM given to \texttt{DAOStarFinder} is taken from the ZTF image header. After \texttt{DAOStarFinder} runs, only sources that fall within the convex hull are accepted as candidates for the asteroid.

The result is a catalog of potential asteroid detections. For image $m$, the pixel coordinates of sources detected in the search region are denoted
\begin{equation*}
X^m=\{x^m_1, \, x^m_2, \, \dots \}.
\end{equation*}
We also compute the overall density of detected sources $\rho_m$ by dividing the number of sources detected in the image by the number of unmasked image pixels.

\subsection{Candidate refitting and propagation to other images\label{sec:candidaterefitting}}
At this point, for each image we have a catalog $X^m$ consisting of many sources that are consistent with being the asteroid of interest, but at most one of these sources is actually a prediscovery detection. The rest are false positives. We tackle the false positive problem by attempting to link sources in multiple images, together with the postdiscovery data, along a single physical orbit. The logic of this step is that it is unlikely that a spurious orbit comes close to multiple sources by chance.

Our linking procedure works as follows. For the $n^\textrm{th}$ candidate source $x^m_n$ in a source catalog, we hypothesize it to be the asteroid, calling it a trial source. The RA and Dec of the trial source are then appended to the postdiscovery data as a new observation and we perform an updated orbit fit\footnote{These fits with appended prediscovery data use the best-fit orbital elements from our postdiscovery fit as an initial guess.}. We are confident in the astrometry of the survey imaging and source catalog and therefore do not include the Huber factor for the candidate sources in the $\chi^2$. Our overall loss with the appended candidate prediscovery observation is

\begin{equation} 
\Lagr^m_n(\theta) = \Lpost(\theta) + \sum_{j=1}^2 \chi^2\left(x^m_n \mid \theta, \sigma_m\right), \label{eq:loss}
\end{equation}
where $\chi^2\left(x^m_n \mid \theta, \sigma_m\right)$ is defined in Equation~\eqref{eq:chi2factor} and the summation from $j=1$ to~2 indicates there is term for both the RA and Dec of the candidate source $x^m_n$. We set the astrometric uncertainty $\sigma_m$ for the sources in image $m$ to be the standard deviation of the PSF divided by the $\SNR$ threshold used to construct the catalog. This is a simple approximation to the result of \citet{King_1983}.

The optimization of Eq.~$\ref{eq:loss}$ yields the best-fit orbital elements $\hat\theta^m_n$ and corresponding orbital element covariance $\hat\Sigma^m_n$.

We next predict the asteroid's position in the other images under the assumption that $\hat\theta^m_n$ is the true orbit. The propagation of the orbit to image $m'$ (called the destination image) gives predicted pixel coordinates
\begin{equation}
\hat x^{m m'}_n = f(\hat\theta^m_n, t_{m'}). \label{eq:refitpred}
\end{equation}
The predicted positions will next be probabilistically matched with the source catalogs of the destination images to reject false positive candidate sources and identify prediscoveries.

\subsection{Linking and false positive rejection\label{sec:linking}}
Once the refitting and propagation process has been iterated through every trial source from each image catalog, the next step is to eliminate the many false positive catalog sources that are unrelated to the asteroid. This is done by testing whether refitted orbits come closer to sources across multiple images than would be predicted by random chance.

This procedure will work best when a candidate source extends the observational arc significantly compared to the postdiscovery data. In this case, the uncertainty $\hat\Sigma^m_n$ on the orbit is drastically reduced compared to $\Sigmapost$, and the asteroid's predicted location in other images is very narrowly constrained. It is unlikely that a spurious source will appear in such a narrow region. In the opposite scenario, a candidate source might lead to a refitted orbit that is still consistent with a large region in other images. This situation is more likely to arise when the candidate image is relatively close in time to the discovery, meaning that a prediscovery in such an image is less impactful from a planetary defense perspective. In the case where there are candidate images both close and far in time from the MPC data, trial orbits for sources in the far images can usefully match with sources in the images nearer in time to the MPC data, but the opposite is not true. 

How exactly do we map out the uncertainty region consistent with a trial orbit? We could once again adopt a Monte Carlo method and sample orbits from the refit covariance $\hat\Sigma^m_n$, propagate these to another image, draw a contour containing them, and define a consistent detection as a catalog source falling within this contour. However, this becomes combinatorially expensive when there are numerous candidate images. We proceed under two assumptions: (1) trial orbits originating from the same image $m$ have similarly-shaped uncertainty regions when propagated to the same destination image $m'$, or equivalently, the geometry of the problem is relatively consistent within the small region of sky that images occupy; and (2) the uncertainty (i.e., standard error) in the predicted location $\hat x^{mm'}_n$ can be described by a 2-d Gaussian distribution. In other words, given the postdiscovery observations augmented with the candidate source $x^m_n$, the asteroid's location in image $m'$ is described by a Gaussian with mean $\hat x^{mm'}_n$ and a $2\times2$ covariance matrix $\hat\Sigma^{mm'}$ that is the same for all sources $n$ in image $m$. The details for estimating $\hat\Sigma^{mm'}$ are given in Appendix~\ref{sec:covarianceestimate}.

Next, we define our probabilistic linking procedure. The derivation that follows gives a measure for how unlikely it is that random sources would align with the orbit’s predicted path across multiple images. 

We construct a likelihood ratio for each candidate source $x^m_n$. The null hypothesis is that the source is a random interloper, image artifact, or noise fluctuation not related to the asteroid, while the alternative is that $x^m_n$ is a real prediscovery detection.

If $x^m_n$ is the true asteroid, then we expect to have a detection in image $m'$ at a position $x$ distributed as a 2-d Gaussian
\begin{equation}
    x \sim N\left(\hat x^{mm'}_n,  \,\, S^{mm'} \right), \label{eq:H1predictedposition}
\end{equation}
with mean $\hat x^{mm'}_n$ and covariance
\begin{equation}
    S^{mm'} \equiv \hat\Sigma^{mm'} + (\sigma_{m'}^2) I_2. \label{eq:sumcov}
\end{equation}
The covariance has two terms. The first reflects the uncertainty in the orbit $\hat\theta^m_n$ and the second is the centroid error of source detection in image $m'$. The latter is the astrometric uncertainty $\sigma_{m'}$ discussed after Equation~\eqref{eq:loss} (we use the same symbol  $\sigma_{m'}$ for centroid error in pixel and sky coordinates; $I_2$ is the $2\times2$ identity matrix).

Motivated by Equation~\eqref{eq:H1predictedposition}, we construct a hypothesis test based on the closest source $x^{m'}_{n'}$ to $\hat x^{mm'}_n$ according to the Mahalanobis distance metric \citep{Mahalanobis1936} defined by the covariance $S^{mm'}$:
\begin{multline}
    r^{mm'}_n = \min_{n'} \left[ \left( (x^{m'}_{n'} - \hat x^{mm'}_n \right)^T (S^{mm'})^{-1} \right. \\
    \left. \vphantom{\left( (x^{m'}_{n'} - \hat x^{mm'}_n \right)^T}
    \left( (x^{m'}_{n'} - \hat x^{mm'}_n \right)
    \right]^{1/2}. \label{eq:closestsourcedef}
\end{multline}

The likelihood ratio is the ratio of probability densities for $r^{mm'}_n$ under the alternative and null hypotheses. The details are worked out in Appendix~\ref{sec:closestsourcepdfs}. The contribution to the log-likelihood ratio from image $m'$ for the candidate source $n$ in image $m$ is (Equation~\eqref{eq:loglikeratioterm_app})
\begin{align}
    \lambda^{mm'}_n(r)
    &= \log \left[ (1-\alpha) + \alpha \left( 1 + \frac{1}{2 \pi \trho_{m'}} \right) e^{-\frac12 r^2} \right].
    \label{eq:loglikeratioterm}
\end{align}
In this equation, $r$ is the smallest Mahalanobis distance from a source in image $m'$ to the asteroid's predicted location $\hat x^{mm'}_n$, using the covariance matrix $S^{mm'}$ (Equations~\eqref{eq:sumcov} and~\eqref{eq:closestsourcedef}). The density $\trho_{m'} = \rho_{m'} \lvert S^{mm'} \rvert^{1/2}$ is the density $\rho_{m'}$ of random sources in image $m'$ (defined at the end of Section~\ref{sec:source_detection}) transformed into isotropic coordinates by $S^{mm'}$ ($\lvert S^{mm'} \rvert$ is the determinant of $S^{mm'}$; Equation~\eqref{eq:transformeddensity}). Finally, $\alpha$ is a prior probability that the asteroid is detected in image $m'$. We found good results by simply setting $\alpha=0.5$, but this is a hyperparameter that may be worth tuning.

In the case where the asteroid is detected in the image $m'$, the Mahalanobis distance will be $r \approx 1$. If the background density of sources in this image is low (i.e., $2\pi \trho_{m'} \ll 1$) the contribution to the log-likelihood ratio is large because it was unlikely to obtain such a small $r$ by chance. On the other hand, if $x^m_n$ were a false positive source (or if the asteroid is not detected in image $m'$, or it is detected but the positional uncertainty is large, or the density of background sources in $m'$ is large) we will have $r^2 \approx 1/\trho_{m'}$ and the contribution to the log-likelihood will be $\approx \log(1-\alpha) < 0$.

The total log likelihood for the candidate source $x^m_n$ is
\begin{equation}
    \lambda^m_n = \sum\limits_{m' \neq m} \lambda^{mm'}_n (r^{mm'}_n ),
    \label{eq:totallikeratio}
\end{equation}
where the sum is over all images $m'$ (other than $m$) intersected by orbit $\hat\theta^m_n$. Large values of $\lambda^m_n$ are evidence of a prediscovery. To determine what counts as large, we standardize the total log-likelihood ratio by subtracting its mean and dividing by its standard deviation under the null hypothesis as described in Appendix~\ref{sec:nulldist}. The result is a standardized log-likelihood ratio (Equation~\eqref{eq:standardizedloglikeratio}), which should be interpreted as prediscovery significance in units of standard deviations above the null hypothesis expectation. A simple threshold like significance $>10$ easily separates prediscoveries from spurious sources (see Figure~\ref{fig:dg1_hist} below). A more precise way to calculate statistical significance is to generate large samples of $\lambda^m_n$ under the null hypothesis (using Equation~\eqref{eq:loglikeuniform}) and directly count the fraction that are greater than the observed $\lambda^m_n$. Alternatively, the samples can be used to fit the parameters of an extreme value or Pareto distribution that describes the false positive tail, accounting for the multiple hypotheses tested \citep[e.g.,][]{Coles2001Introduction}.

\subsection{Self-consistency of significant candidates\label{sec:selfconsistency}}
The result of the linking procedure is a collection of significant candidate sources whose best-fitting orbits pass improbably close to sources in other images. A last check is to make sure these orbits all represent the same asteroid trajectory. There are multiple ways this might be done. A simple approach is to keep track of the identity of the closest source $n'$ to the predicted position in each image $m'$, i.e., save the value of $n'$ that minimized the distance in Equation~\eqref{eq:closestsourcedef}. For each significant candidate, we require that this list of closest sources contain the other candidates. Another approach would be to fit an orbit to the postdiscovery data along with all significant candidate prediscoveries and require an acceptable goodness of fit.

\section{Results\label{sec:results}}

In this section, we consider the prospects for making long-arc extension prediscoveries using the existing ZTF archive, demonstrate successful prediscoveries using our pipeline for two NEAs, and perform null tests on asteroids that are too faint to be seen in ZTF.

\subsection{Candidates for prediscovery analysis\label{sec:predisco_scanning}}

To search for objects with useful and potentially achievable prediscoveries, we query the list of NEOs from the Jet Propulsion Laboratory Small Body Database (SBDB)\footnote{\url{https://ssd.jpl.nasa.gov/tools/sbdb_lookup.html}. Query date 2025-09-25.} and retain objects with a discovery date after the beginning of the ZTF survey. Using the orbital elements provided by SBDB, we identify candidate ZTF image intersections as in Section~\ref{sec:source_detection}, except that we only propagate the best-fitting orbit $\thetapost$ rather than a sample of orbits from its covariance. Additionally, we calculate a potential arc extension ratio, defined as the current observational arc divided by the arc extension that would be achieved if the earliest candidate image contained a prediscovery. We deem an object to have a potential prediscovery if there are at least five candidate intersections, meaning our pipeline has sufficient images to link detections across for false positive rejection.

\begin{figure}
    \centering
    \includegraphics[width=\linewidth]{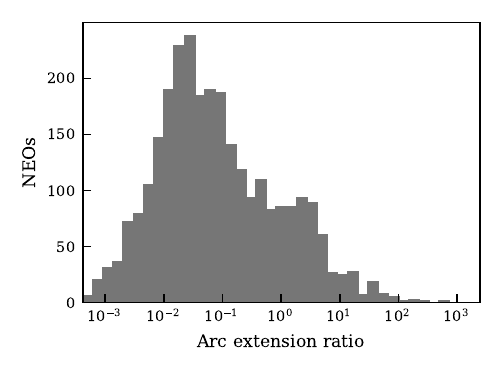}
    \caption{Potential arc extension ratios of 2{,}676 NEAs queried from JPL SBDB with at least 5 candidate images in ZTF. Arc extension ratio is defined as the potential increase in arc divided by the current arc.\label{fig:arc_ext}}
\end{figure}
Figure~\ref{fig:arc_ext} shows the potential of automated prediscovery using ZTF. Out of the 18{,}808 NEOs discovered after the start of ZTF, 2676 are expected to be present in $\geq 5$ ZTF images at magnitudes bright enough to make it into low-significance image catalogs (the figure is essentially unchanged if the catalog threshold is raised from $3\sigma$ to $4\sigma$). Of these, the figure shows the potential arc extension ratios that will be achieved by successful prediscovery. There are 490 NEOs for which we can expect to at least double their arcs using already-collected data. 

Our method is survey-agnostic, so this number of potentially substantial arc extensions in a single survey demonstrates the potential for a method such as the one proposed here, especially as future surveys detect substantially more NEOs in the coming years. The availability of archival wide-field survey images years in advance offers the potential to immediately improve initial orbit determination for detections made with the new surveys coming online. 

\subsection{Prediscovery of 2021 DG1}

\begin{figure}
    \centering
    \includegraphics[width=1\linewidth]{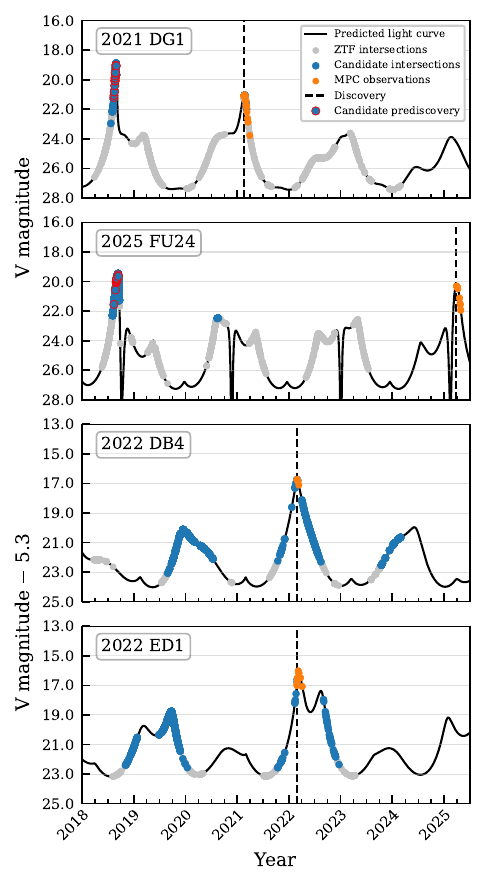}
    \caption{Predicted light curves of selected NEAs based on their postdiscovery orbit fits. Gray points show all ZTF images that should contain the asteroid. Candidate intersections (blue) are ZTF images in which the asteroid is predicted to be brighter than the image's $3\sigma$ detection threshold. The presence of candidate intersections before the discovery date allows for significant arc extension. The points circled in red in the top two panels mark successful prediscovery detections in ZTF. They correspond to candidate sources with a likelihood significance greater than $10\sigma$ (see Figure~\ref{fig:dg1_hist}). The bottom two panels demonstrate null detections: we artificially brighten the reported magnitudes for 2022~DB4 and 2022~ED1 by 5.3~mag but they are in fact undetectable in ZTF.}
    \label{fig:dg1_lightcurve}
\end{figure}

The potentially hazardous asteroid (PHA) 2021~DG1 was discovered by Pan-STARRS in February 2021. Its absolute magnitude is $H=22$, and it has a semimajor axis of $a=1.65~\mathrm{au}$ and an Earth MOID of 0.0026~au. With an observational arc length of just 42~days, the object is attributed an uncertainty parameter $\mathrm{U}=6$ by the MPC. Its light curve, shown in Figure~\ref{fig:dg1_lightcurve}, indicates that numerous ZTF images could extend the observation arc by $\sim$2.5~yr. The asteroid has close approaches with Earth three times in the next century, including in 2107 when NEODyS\footnote{\url{https://newton.spacedys.com/index.php}} reports a minimum close approach distance $< 0.001\,\mathrm{au}$. 

\begin{figure}
    \centering
    \includegraphics{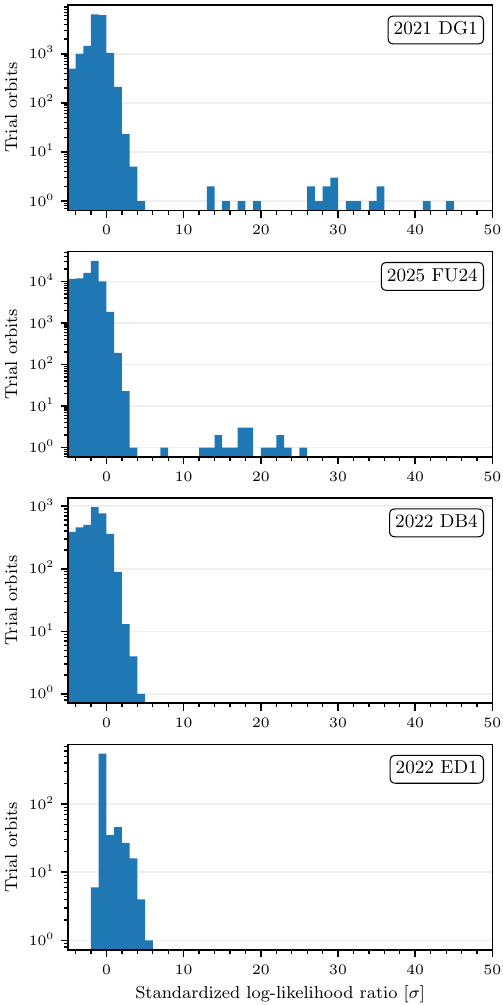}
    \caption{Prediscovery significance of candidate sources obtained from our pipeline for selected NEAs. Each panel shows a histogram of the standardized log-likelihood ratio (in units of standard deviations above the null hypothesis expectation) for orbits corresponding to trial sources detected in search regions. There is clear evidence of prediscovery for 2021~DG1 and 2025~FU24, where orbits successfully predict the asteroid's position across multiple images. The bulk of orbits with standardized log-likelihoods near 0 reflect the null distribution corresponding to spurious sources unrelated to the asteroids (orbits with significance less than $-5$ are not shown). The asteroids 2022~DB4 and 2022~ED4 are too faint to be detected in ZTF.
    \label{fig:dg1_hist}}
\end{figure}

Our prediscovery pipeline identified 879 candidate images and 18{,}394 total sources in the resulting search regions, which cover $11.7\,\deg^2$ of total sky area at ZTF epochs. Each source results in a trial orbit after appending its astrometric position to the MPC data and refitting. Standardized log-likelihood ratios are calculated and the results are shown in Figure~\ref{fig:dg1_hist}. Orbits with a large positive standardized log-likelihood ratio are confirmed to comprise orbits originating from true detections of 2021~DG1, as seen by examining individual images as in Figure \ref{fig:ztfimage}. The 20 candidate sources with log-likelihood significance above $10\sigma$ correspond to orbits that intersect each other (see Section~\ref{sec:selfconsistency}). For the highest log-likelihood ratio orbit, there are 19 images with a detection within 1'' of the predicted location.

\begin{figure}
    \centering
    \includegraphics[width=1\linewidth]{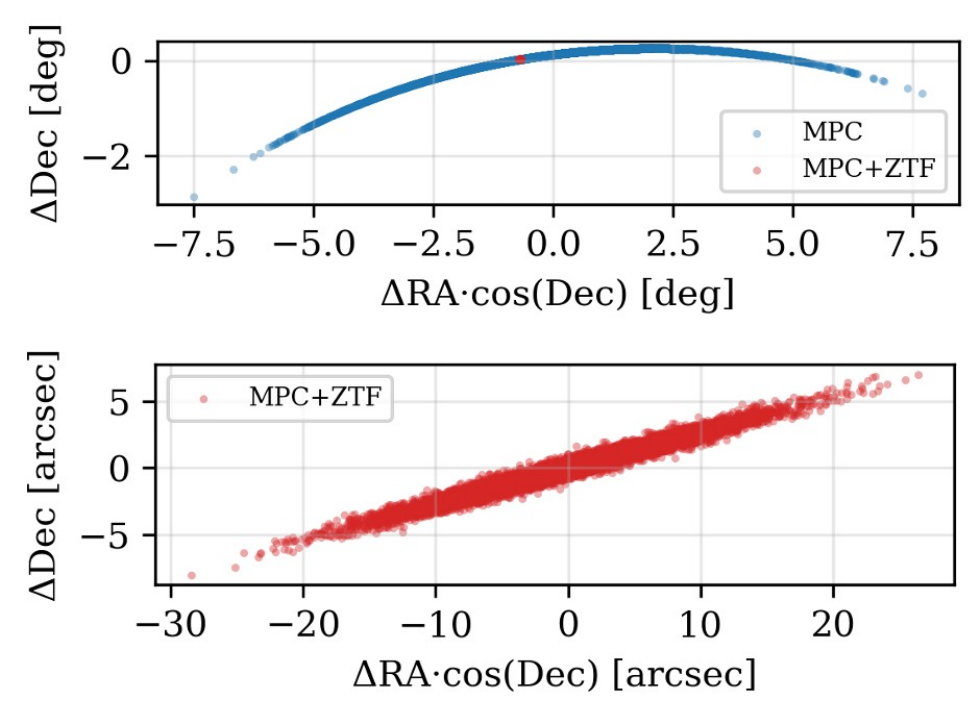}
    \caption{Upper panel: Sky plane error for 2021~DG1 on 2035-07-31, which corresponds to the first time it is brighter than $\mathrm{V}=23$. Fits of its orbit with (red) and without (blue) ZTF prediscoveries were propagated with a Keplerian propagator. Lower panel: zoom in on the sky-plane error for the combined fit.}
    \label{fig:sky_error}
\end{figure}
In Figure \ref{fig:sky_error}, we show the practical effect of extending 2021~DG1's arc from a few weeks to $\sim$2.5~yr. 2021~DG1 will remain difficult to observe until 2035, which is the next time it becomes brighter than V$=23$. We propagated samples from the orbit covariance of fits to the MPC data alone and from a fit to the combined MPC and ZTF data to 2025-07-31. The sky-plane error is smaller by orders of magnitude and can be easily followed up 10~years from now without requiring a search.

\subsection{Prediscovery of 2025 FU24}
The earliest known image of 2025~FU24, a recently
discovered $H=21.9$ NEO, is from Pan-STARRS in March 2025. It has an observational arc of 31~days. Its Earth MOID is 0.09~au. Figure \ref{fig:dg1_lightcurve} shows candidate intersections with ZTF in 2018. 

A significant difference from the case of 2021~DG1 is the much longer potential arc extension, as 2025~FU24 was discovered 4~yr later. This manifests as a significantly higher computational cost --- projecting the postdiscovery uncertainty further backwards in time expands the sky regions we must search and thus the number of trial orbits. Since ZTF image files are stored as CCD quadrant files, the orbit covariance intersects many image files across each ZTF exposure. Our prediscovery pipeline identified 5048 candidate image intersections and 184{,}452 sources within image search regions that cumulatively span $113.6\,\deg^2$ of ZTF imagery. A manual prediscovery search of such a sky area would be daunting. Nonetheless, the pipeline is successful and the orbit with the highest significance intersects 18 catalog sources within 1''. Prediscovery of this object increases its observational arc by a factor of 78. Neither 2025~FU24 nor 2021~DG1 has any previous reported MPC observations from ZTF.

\subsection{Examples of null detections}

In this subsection, we demonstrate the output of our pipeline in the absence of a detectable asteroid. To do this, we examined two asteroids far too faint to be detected in the ZTF imagery prior to the earliest observation in the MPC data.

2022~DB4 is an $H=24.9$ NEO discovered by Pan-STARRS in February 2022. It has an uncertainty parameter $\textrm{U}=7$ and an observational arc of just 13~days. 2022~DB4 was manually chosen by the procedure described in Section~\ref{sec:predisco_scanning} to have no valid candidate images while still having a peak in its light curve a few years prior to its discovery, allowing for a potential arc extension if the object was brighter. We then artificially brightened the apparent magnitude for the object by 5.3~mag so that the pipeline included a similar number of total candidate images as in the prior examples. 

Our prediscovery pipeline identified 833 candidate image intersections and 4943 sources within those images. In contrast to the successful prediscoveries of 2021~DG1 and 2025~FU24, all trial orbits exhibit a standardized log-likelihood ratio close to zero with no outliers in the histogram (see Figure~\ref{fig:dg1_hist}). We manually confirmed the orbits with the highest log-likelihood ratio to be false detections of difference image artifacts. These orbits are not self-consistently aligned with the sources in other images. 

A similar procedure was carried out for 2022~ED1, an $H=24.5$ NEA with an uncertainty parameter of $\textrm{U}=6$ and an observational arc of 38~days. Our pipeline returned a similar standardized log-likelihood ratio distribution and a similar finding of no self-consistent detections aligned within the ZTF images. This object intersected 812 images with 694 sources detected in the search regions. 

\section{Discussion}\label{sec:conclusions}
In this section we summarize our results and discuss future work to extend our pipeline's capabilities. 

We have introduced an automated, probabilistic method for asteroid predetection and demonstrated it using archival ZTF difference images. We discovered 19 and 18 prediscovery detections for NEOs 2021~DG1 and 2025~FU24, respectively, which decrease orbital uncertainty by extending observational arcs by years back from the first reported MPC observations. 

Our method is capable of processing thousands of survey images and linking hundreds of thousands of trial orbits based on low-significance source catalogs generated from survey images. With 6~yr of ZTF data on disk, we demonstrate a capability to make prediscovery observations that can extend arcs by up to factors of 78 (in the case of 2025 FU24). Our method requires access to the images because catalogs are generated at a lower detection threshold than surveys typically do. This results in many false positives, which we rule out by requiring self-consistent detections across multiple survey images. We handle this linking problem in a probabilistic formalism, comparing against the null hypothesis that the image sources are randomly distributed and not associated to a real object that can be linked to the MPC data along a physical orbit. Our method is survey-agnostic and could be applied to other wide-field surveys like Pan-STARRS or Rubin. It is equally suited to be used for prediscovery within a survey itself, as an automated post-processing step after standard linking algorithms like HelioLinc3D \citep{heliolinc,heliolinc3d}) determine an initial orbit.

\subsection{Future work}
Future work will address scalability to enhance the handling of more uncertain orbits over longer time periods. The required computation time of our pipeline scales with the uncertainty of the asteroid. More uncertain orbits have larger search regions in survey images. This increases the number of images, the size of source catalogs, and the number of trial orbits that need to be processed. We demonstrated success on objects with an uncertainty parameter $\mathrm{U}=6$ and delivered arc extensions nearly 100 times larger than the current observational arc. We believe that our current code can handle NEAs with $\textrm{U} \leq 7$, with more uncertain orbits requiring better parallelization.

In the case of 2021~DG1, some prediscovery observations exhibited trailing losses. These lowered our detection sensitivity since we ran \texttt{DAOStarFinder} using a symmetric Gaussian kernel. Our method could be improved to include a signal-matched filter based on the exposure time, proper motion, and seeing of each image. The orientation and length of the streak can be predicted from the postdiscovery data so incorporating a streak kernel would enable fainter prediscoveries to be made for trailing objects.

Beyond allowing for fainter detections, trailing can be a powerful discriminator between the true asteroid and spurious sources or image artifacts. For sources that trail (e.g., because of close approaches or higher resolution instruments), the vast majority of spurious detections will not match the predicted trail length and orientation. In the source detection step (Section~\ref{sec:source_detection}), instead of just measuring the position of each candidate source, one would fit a trailed model and promote $x^m_n$ to a four-dimensional quantity that includes streak length in the horizontal and vertical directions. To a good approximation, a source's position in a search region will be independent of its trail vector, which means the log-likelihood statistic (Equation~\eqref{eq:loglikeratioterm_app}) would pick up a simple additive term that measures the discrepancy between the predicted and measured trail.

Figure~\ref{fig:dg1_lightcurve} shows that the window in which NEA prediscoveries are possible is often short. Prediscovery observations occurred over just a short period of time for both 2021~DG1 and 2025~FU24. However, stacking methods could enhance the capability to make detections of fainter objects that are not bright enough to appear in catalogs. Digital tracking along nonlinear motion is generally too computationally expensive for blind searches of NEAs \citep{2025A&C....5300987G, Alex}, but targeted searches seeded on fits to MPC data and low-significance catalogs could enable the prediscovery of fainter objects than we demonstrated here \citep[e.g.][]{2025arXiv250926279S}.  Along these lines, we noticed in our demonstrated cases that even fainter prediscoveries can be made through visual inspection of image cutouts centered on the predicted locations for significant trial orbits, even when no catalog source was detected nearby. This suggests it may be possible to perform a ``second pass'' at even lower catalog thresholds near promising orbits.

\subsection{Outlook on prediscovery efforts}
Prediscovery searches have often been handled in an ad hoc manner. Recent examples such as the case of 2024~YR4, where a nonzero impact risk was identified, demonstrated in real time the importance of having ready-to-go software for this task. There was an international effort to search for prediscovery detections in archival databases from telescopes around the world, but the difficulty in handling the long Line of Variations \citep[LOV;][]{milani2005nonlinear} from the fit led us to imagine a more probabilistic approach. We were not alone in this assessment, as some sought to use null detections in archival images to rule out portions of the LOV and inform impact probability assessments. In the end, 2024~YR4 was not found in archival images from survey instruments, so this methodology is not guaranteed to work for any given asteroid. However, in the next year, Rubin will cause a step function in the discovery rate of NEOs. More cases like 2024~YR4 are possible, and perhaps even likely, given the extreme depth of Rubin images. The coming era of Rubin, with NEO Surveyor and NEOMIR to follow, requires a methodology to handle these cases, and asteroid prediscovery pipelines need access to archival survey imagery to be most useful.

\begin{acknowledgments}
The authors thank Peter McGill and Lila Braff for their input over the course of this study. 

S. Li acknowledges support from the LLNL Space Science Institute Summer Internship Program. 

This work was performed under the auspices of the U.S. Department of Energy by Lawrence Livermore National Laboratory under Contract DE-AC52-07NA27344. 

This work was supported by the Lawrence Livermore National Laboratory LDRD Program under Project 2023-ERD-044. The LLNL document number is LLNL-JRNL-2011856. 

This work is based on observations obtained with the Samuel Oschin Telescope 48-inch and the 60-inch Telescope at the Palomar Observatory as part of the Zwicky Transient Facility project. ZTF is supported by the National Science Foundation under Grants No. AST-1440341 and AST-2034437 and a collaboration including current partners Caltech, IPAC, the Oskar Klein Center at Stockholm University, the University of Maryland, University of California, Berkeley, the University of Wisconsin at Milwaukee, University of Warwick, Ruhr University, Cornell University, Northwestern University and Drexel University. Operations are conducted by COO, IPAC, and UW.
\end{acknowledgments}

\software{Astropy \citep{2013A&A...558A..33A,2018AJ....156..123A,2022ApJ...935..167A},
          DAOPHOT \citep{1987PASP...99..191S},
          Matplotlib \citep{Hunter:2007},
          NumPy \citep{2020NumPy-Array},
          SciPy \citep{2020SciPy-NMeth}.
          }

\newpage

\appendix

\section{Estimation of image-space covariance\label{sec:covarianceestimate}}

The image-space covariance induced by the covariance in the orbital elements can be mapped via a fast, first-order approximation method. The ephemeris function $f(\theta, t)$ takes a set of 6-d orbital elements $\theta$ and returns the predicted 2-d pixel coordinates $x$ in an image taken at time $t$. The distribution of orbital elements is described by a mean $\theta_0$ and covariance $\Sigma_\theta$. Expanding to first order around $\theta_0$,
\begin{equation*}
    x = x_0 + J \, (\theta - \theta_0) + \mathcal{O}\left((\theta-\theta_0)^2\right),
\end{equation*}
where $x_0 = f(\theta_0, t)$ and $J=\partial x/\partial\theta$  (a $2\times6$ matrix) is the Jacobian of $f$ evaluated at $\theta_0$. The covariance of $x$ is
\begin{equation}
    \Sigma_x = J \Sigma_\theta J^T. \label{eq:JsigJ}
\end{equation}
Rewriting the positive definite $\Sigma_\theta$ as $QQ^T$ (e.g. by Cholesky decomposition or diagonalization), Equation~\eqref{eq:JsigJ} becomes
\begin{equation}
    \Sigma_x = (JQ)(JQ)^T. \label{eq:covdecomp}
\end{equation}
Each column of $JQ$ is the Jacobian multiplied by a column of $Q$, which is the approximate change in pixel coordinates when the orbital elements shift by the column of $Q$. This suggests a simple recipe for calculating $JQ$ by perturbing the orbital parameters by $\delta\theta_i \propto Q_i$ and recalculating the ephemeris. The $i^\mathrm{th}$ column of $JQ$ can be approximated by the central finite difference

\begin{equation}
    (JQ)_i \approx \frac{f(\theta_0 + h Q_i, t) - f(\theta_0 - h Q_i, t)}{2h},
    \label{eq:JQest}
\end{equation}
where $Q_i$ is a column of $Q$ and $h$ is a small scalar step size. The approximation becomes exact as $h \to 0$.

This procedure, as opposed to sampling potentially thousands of orbits and thus performing thousands of integrations, only requires 12 orbit integrations to obtain $\Sigma_x$, drastically reducing computational cost.

In our implementation, $Q$ is based on the eigendecomposition $\Sigma_\theta = O D O^T$, where $O$ is orthogonal and $D$ is diagonal with positive elements, so that $Q = O D^{1/2}$. The step size $h$ is initially set to 1, and is reduced iteratively if the shift in pixel coordinates (numerator of Equation~\eqref{eq:JQest}) is greater than 10 pixels. In practice, the initial step size is usually sufficient for asteroids of interest.

In the linking step (Section~\ref{sec:linking}), we need to estimate $\hat\Sigma^{mm'}$, the position uncertainty in image $m'$ of an orbit fitted to a candidate source in image $m$. We do this by picking an exemplar source $x^m_n$ in image $m$ whose best-fit orbit lands within image $m'$ (i.e., $\hat x^{mm'}_n$ is within the boundaries of $m'$) and computing the above covariance $\Sigma_x$ (Equation~\eqref{eq:covdecomp}) for $\theta_0=\hat\theta^m_n$ and $\Sigma_\theta = \hat\Sigma^m_n$. The resulting covariance is the $\hat\Sigma^{mm'}$ of Section~\ref{sec:linking}, which is used for all sources $n$ in image $m$ whose best-fit orbit intersects image $m'$.

\section{Log-likelihood ratio test statistic\label{sec:closestsourcepdfs}}

For candidate source $n$ in image $m$, evidence of whether it represents a true prediscovery is based on the observed values of $r^{mm'}_n$ (Equation~\eqref{eq:closestsourcedef}) in each destination image $m'$. Considering a single destination image $m'$, and dropping all sub and superscripts for clarity, $r$ is the distance from the asteroid's predicted position in $m'$ to the closest source in that image as measured by the Mahalanobis distance with covariance $S$ (Equation~\eqref{eq:sumcov}). This appendix derives the probability distributions for $r$ under the null and alternative hypotheses.

The null hypothesis assumes that the trial orbit is based on a false detection in image $m$. Therefore, the orbit's location in image $m'$ is not correlated with any of the sources detected in that image. We model the spatial distribution of sources in image $m'$ as a 2-d point process with constant density $\rho$ ($\rho_{m'}$ in the notation of Section~\ref{sec:source_detection}). This is an oversimplification since detected sources are often clumped up around an image artifact (a poorly subtracted star or diffraction spike, for instance), but it is good to a first approximation.

The Mahalanobis distance becomes a Euclidean distance via the linear change in coordinates from image coordinates $x$ to $z = L^{-1} x$, where the covariance is decomposed as $S = L L^T$ (e.g. by Cholesky decomposition). In $z$-coordinates, $r$ is just the Euclidean distance to the closest image source.

A homogeneous point process remains homogeneous under linear transformation, just with a scaled density $\rho \to \trho$,
\begin{equation}
    \trho = \rho \, \vert S\vert^{1/2}, \label{eq:transformeddensity}
\end{equation}
where $\vert S \vert$ is the determinant of $S$. Therefore, under the null hypothesis, $r$ is described by the probability density function (pdf)
\begin{equation}
    f_0(r) = e^{-\trho \pi r^2} \, 2\pi \trho \,r, \label{eq:H0pdf}
\end{equation}
i.e., the probability that the closest source is between $r$ and $r+dr$ is the probability that there are no sources within $r$ times the probability that there is a source between $r$ and $r+dr$. This is the pdf of a Rayleigh distribution with scale parameter $1/\sqrt{2 \pi \trho}$.

Under the alternative hypothesis, the trial orbit is the actual orbit of the asteroid. Then, in image $m'$ we expect to find a source (the actual asteroid) whose position is distributed as a 2D Gaussian with covariance $S$ centered on the predicted position (see Equation~\eqref{eq:H1predictedposition}). In the isotropic $z$-coordinates, this is just a 2-d Gaussian density with diagonal, unit covariance so that Mahalanobis distance $r$ is distributed as another Rayleigh distribution,
\begin{equation*}
    f_\mathrm{ast}(r) = e^{-\frac12 r^2}\, r. \label{eq:astpdf}
\end{equation*}

The alternative hypothesis is slightly more complicated because of two factors. First, the asteroid may not be detected in image $m'$. We are working with images where, by choice, the asteroid is often at very low flux. Further, uncertainties in the light curve due to tumbling may push the asteroid below the threshold in a particular image. To account for this, we introduce a prior probability $\alpha$ that the asteroid will make it into the catalog for image $m'$. Our method is likely not particularly sensitive to the value of $\alpha$ as long as it is not set to $1$. If $\alpha=1$ and the asteroid is not detected in the image, we may find $r \gg 1$, which would severely disfavor the alternative hypothesis (see Equation~\eqref{eq:loglikeratioterm_app} with $\alpha=1$). We found good results by setting this hyperparameter to $\alpha=1/2$.

The second subtlety in the alternative hypothesis is that even in the case when the trial orbit is the asteroid's true orbit, it may be that the closest source in image $m'$ is a random catalog source unrelated to the asteroid. This may happen if the density of sources in image $m'$ is very high and/or if the uncertainty in the fitted orbit $\hat\Sigma^m_n$ is large, so that the asteroid is likely to be found far from the best-fit prediction $\hat x^{mm'}_n$.

Accounting for these effects, the pdf of $r$ under the alternative hypothesis is
\begin{equation}
    f_1(r) = (1-\alpha) f_0(r) + \alpha\left[f_\mathrm{ast}(r) e^{-\trho \pi r^2} + f_0(r) e^{-\frac12 r^2} \right]. \label{eq:H1pdf}
\end{equation}
The first term represents the case that the asteroid is not detected in the image. The first term in brackets is the probability that the asteroid is detected at Mahalanobis distance $r$ and no catalog sources are detected within $r$, while the second term is the opposite situation (the closest catalog source detected at $r$ with the asteroid detected beyond $r$).

Combining Equations~\eqref{eq:H0pdf} and~\eqref{eq:H1pdf}, the log-likelihood ratio for the image is (see Equation~\eqref{eq:loglikeratioterm})
\begin{equation}
    \lambda(r) \equiv \log \frac{f_1(r)}{f_0(r)} = \log\left[(1-\alpha) + \alpha \left( 1+ \frac{1}{2\pi\trho}\right) e^{-\frac12 r^2} \right]. \label{eq:loglikeratioterm_app}
\end{equation}

\subsection{Distribution under the null hypothesis\label{sec:nulldist}}

It is straightforward to obtain the sampling distribution of the likelihood ratio under the null hypothesis. If the trial orbit is based on a spurious detection, the Mahalanobis distance $r$ in each image will be an independent sample from the null distribution $f_0(r)$ (Equation~\eqref{eq:H0pdf}). The total log-likelihood ratio is then the sum of many independent random variables, each governed by the two parameters $\trho$ and $\alpha$. The sampling distribution can be used to give a precise detection significance, or $p$ value, for the trial orbit, i.e., the probability, under the null hypothesis, of obtaining a log-likelihood ratio greater than observed.

It is possible to compute the $p$ value, either by Monte Carlo sampling of an $r$ value for each image or by convolving the null $\lambda$-distributions for each image together via Fourier methods \citep[e.g.][Appendix~A.2]{2015PhRvD..91h3535G}. Here, we provide a simple recipe to standardize the log-likelihood ratio by subtracting its mean under the null hypothesis and dividing by its standard deviation. This produces a significance in ``sigma units'', e.g. a standardized log-likelihood ratio of 10 means the orbit under consideration is a ``10 sigma'' outlier under the null hypothesis.

The cumulative distribution function (cdf) of $r$ under the null hypothesis is the integral of Equation~\eqref{eq:H0pdf},
\begin{equation*}
    F_0(r) = 1 - e^{-\trho \pi r^2},
\end{equation*}
which means that Equation~\eqref{eq:loglikeratioterm_app} can be rewritten as
\begin{equation}
    \lambda = \log\left[(1-\alpha) + \alpha\left(1 + \frac{1}{2\pi\trho}\right)\left(1 - F_0\right)^{1/(2\pi\trho)}\right].
    \label{eq:loglikeuniform}
\end{equation}
Under the null hypothesis, the cdf $F_0(r)$ is distributed as a uniform random variable between 0 and 1. Simple numerical integration of Equation~\eqref{eq:loglikeuniform} (and its square) from $F_0=0$ to~$1$ gives the mean $\mathrm{E}_\lambda$ and variance $\mathrm{Var}_\lambda$ of $\lambda$ under the null hypothesis. It is easy to precompute a lookup table of $\mathrm{E}_\lambda(\trho)$ and $\mathrm{Var}_\lambda(\trho)$ for a range of $\trho$. Then for a given trial orbit, which intersects images $m'$, one interpolates the lookup table for the values of $\trho_{m'}$ and computes the standardized log-likelihood ratio:
\begin{equation}
    \mathrm{Signif} = \dfrac{\displaystyle\sum\limits_{m'}\lambda(r^{m'}) - \sum\limits_{m'} \mathrm{E}_\lambda(\trho_{m'})}{\displaystyle\sqrt{\sum\limits_{m'}\mathrm{Var}_\lambda(\trho_{m'})}},
    \label{eq:standardizedloglikeratio}
\end{equation}
where the first term in the numerator is the log-likelihood ratio for the observed Mahalanobis distances $r^{m'}$ (see Equation~\eqref{eq:totallikeratio}). Equation~\ref{eq:standardizedloglikeratio} is used to find the prediscovery significance of the trial orbits in our demonstration cases (Figure~\ref{fig:dg1_hist}).

\bibliography{main}{}
\bibliographystyle{aasjournalv7}

\end{document}